\documentclass[sigconf]{acmart}
\usepackage{booktabs}  
\usepackage{makecell}  
\usepackage{amsmath,amsfonts}
\usepackage{algorithmic}
\usepackage{algorithm}
\usepackage{array}
\usepackage{enumitem}
\usepackage[caption=false,font=normalsize,labelfont=sf,textfont=sf]{subfig}
\usepackage{textcomp}
\usepackage{stfloats}
\usepackage{url}
\usepackage{verbatim}
\usepackage{graphicx}
\usepackage{xcolor}
\usepackage{multirow}

\usepackage{lettrine}
\AtBeginDocument{%
  }

\setcopyright{none}
\renewcommand\footnotetextcopyrightpermission[1]{}

\makeatletter
\def\ps@headings{%
  \let\@oddhead\@empty
  \let\@evenhead\@empty
  \def\@oddfoot{\normalfont\hfil\thepage\hfil}%
  \def\@evenfoot{\normalfont\hfil\thepage\hfil}%
}
\let\ps@plain\ps@headings
\def\@acmConference{}
\def\@acmBooktitle{}
\renewcommand\acmConference[4]{}
\makeatother

\begin{document}

\title{Safety and Risk Pathways in Cooperative Generative Multi-Agent Systems: A Telecom Perspective}

\author{Zeinab Nezami, Shehr Bano, Abdelaziz Salama, Maryam Hafeez, Syed Ali Raza Zaidi}
\affiliation{%
  \institution{School of Electrical and Electronic Engineering, University of Leeds}
  \city{Leeds}
  \country{United Kingdom}
}


\renewcommand{\shortauthors}{Nezami.et al.}

\begin{abstract}
   Generative multi-agent systems are rapidly emerging as transformative tools for scalable automation and adaptive decision-making in telecommunications. Despite their promise, these systems introduce novel risks that remain underexplored, particularly when agents operate asynchronously across layered architectures. This paper investigates key safety pathways in telecom-focused Generative Multi-Agent Systems (GMAS), emphasizing risks of miscoordination and semantic drift shaped by persona diversity. We propose a modular safety evaluation framework that integrates agent-level checks on code quality and compliance with system-level safety metrics. Using controlled simulations across 32 persona sets, five questions, and multiple iterative runs, we demonstrate progressive improvements in analyzer penalties and Allocator–Coder consistency, alongside persistent vulnerabilities such as policy drift and variability under specific persona combinations. Our findings provide the first domain-grounded evidence that persona design, coding style, and planning orientation directly influence the stability and safety of telecom GMAS, highlighting both promising mitigation strategies and open risks for future deployment.
\end{abstract}






\maketitle

\section{Introduction}

GMAS offer transformative potential for critical infrastructure by enabling scalable automation, adaptive decision-making, and collaborative intelligence~\cite{frai.2025.1621963}. In telecommunications, these systems promise self-evolving networks~\cite{chaoub2022hybrid,qian2024self,frcmn.2025.dawn}, where agents autonomously perceive, reason, and adjust network behavior in real time. However, GMAS introduce novel classes of risk that differ from those in single-agent or traditional automation frameworks~\cite{hammond2025multi}, which remain largely unexplored. The scarcity of real-world deployments combined with unresolved safety challenges in even single-agent AI, highlights the urgent need to identify and mitigate emergent risks prior to widespread adoption~\cite{hammond2025multi}.

As wireless communication evolves toward self-evolving 6G networks, AI-driven systems will autonomously integrate telemetry, user intent, and environmental signals to refine internal policies, decision logic, and workflow across layers~\cite{chaoub2022hybrid,qian2024self,liang2023code}. Open Radio Access Network (O-RAN)~\cite{airan2025}, together with large language model (LLM)-enhanced intelligence~\cite{mondal2023llms,bariah2024large}, enables dynamic, programmable control, allowing networks to adapt to both operational and societal demands. Yet existing multi-agent evaluation frameworks~\cite{samuel2024personagym,wang2024towards,mao2023alympics} largely emphasize social interaction, persona adherence, or game-theoretic reasoning, while overlooking operational risks in high-stakes domains.

Despite their promise, GMAS are vulnerable to three primary failure modes~\cite{hammond2025multi}: miscoordination, conflict, and collusion. This paper focuses on miscoordination, which arises when agents act on incomplete or inconsistent information, leading to incompatible actions. For example, in an O-RAN deployment, one agent may reserve spectrum based on local telemetry while another follows a conflicting plan derived from a different state, causing resource contention or dropped connections. Such failures emerge naturally from asynchronous actions, divergent tool outputs, or mismatched knowledge contexts, without requiring adversarial intent. Information asymmetry is particularly pronounced in retrieval-augmented generation (RAG) setups, where agents query distinct document stores or maintain non-overlapping memory contexts. Given their layered architecture, real-time constraints, and safety-critical operations, telecom systems provide a compelling testbed for studying these dynamics.

This work proposes a GMAS for network orchestration and automation in O-RAN, and uses it as a testbed to develop and validate a safety evaluation framework for cooperative GMAS. Our framework evaluates agents that share a common objective while executing modular tasks through distributed roles with asymmetric context. 
Our contributions are as follows:
(i) Safety evaluation framework: We design a domain-agnostic framework that combines agent-level checks with system-level safety metrics, and demonstrate its application in O-RAN and RAN Intelligent Controller (RIC) contexts.
(ii) Knowledge–persona integration: We combine heterogeneous knowledge layers (RAG and graph-RAG) with systematic persona variation, and show how these dimensions jointly influence safety, alignment, and reproducibility in multi-agent systems.
(iii) Proposed GMAS for telecom orchestration: We implement a cooperative GMAS for O-RAN automation, providing a concrete instantiation of our evaluation framework.
(iv) Empirical insights: Through multi-run experiments, we uncover progressive improvements in GMAS performance, agent consistency, and embedding stability.
(v) Open resources: We release a GitHub repository containing datasets and code to support reproducibility.

The paper is organised as follows: Section~\ref{sec:rw} reviews related work, Section~\ref{sec:sd} presents the GMAS and safety framework, Section~\ref{sec:eva} reports experiments and results, and Section~\ref{sec:con} concludes with future directions.

\section{Related Work}\label{sec:rw}

\begin{figure*}[!hbpt]
    \centering
    \includegraphics[clip, trim=0cm 15.8cm 1cm 3cm,width=\textwidth]{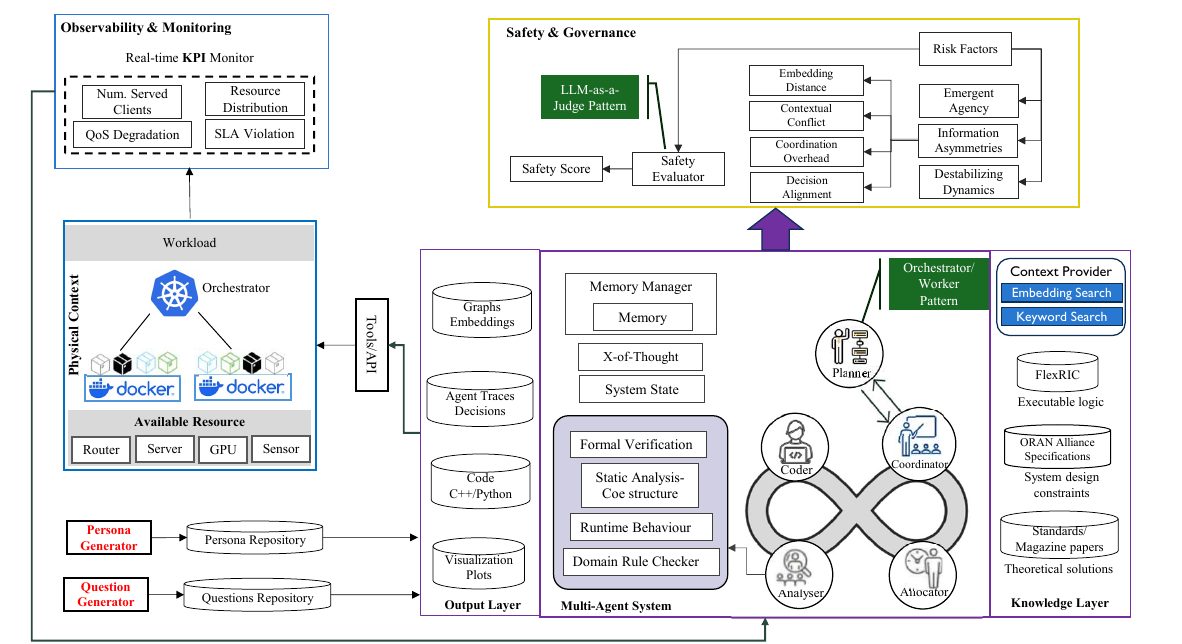}
    \caption{Architecture of safety framework for generative-based multi-agent systems in O-RAN setup}
    \label{fig:architecture}
\end{figure*}
Generative AI (GenAI) has demonstrated strong potential in automating telecom and reducing human effort in tasks such as troubleshooting recommendations~\cite{bosch2022integrating}, code refactoring~\cite{du2023power}, and network configuration~\cite{jiang2023large,xu2024cloudeval}. While recent work has explored GMAS for telecom, the emphasis has largely been on benchmarking~\cite{nezami2025descriptor} and automation~\cite{qin2025generative} rather than safety. To the best of our knowledge, there is no dedicated research addressing safety risks in GMAS for telecom. Related discussions have instead focused either on the safety of GenAI in other domains, or on artificial intelligence safety for telecom more generally. We review three strands of related work: (i) GMAS for telecom, (ii) safety in GenAI-based single-agent systems across domains, and (iii) safety evaluation for GMAS frameworks in telecom.

Panek et al.~\cite{t1} propose a modular framework for cloud-native 5G networks that employs chain-of-thought reasoning before delegating tasks to cloud management tools. Zou et al.~\cite{t2} introduce an edge-based multi-model approach that deploys LLM on wireless devices rather than a central server, reducing reliance on centralized infrastructure but increasing computational overhead. Similarly, Li et al.~\cite{t3} design a multi-agent system for agile and autonomous telecom workflows, where specialized agents handle prompt engineering, fine-tuning, and resource optimization to enhance adaptability. These studies demonstrate the feasibility of GenAI-based multi-agent orchestration in telecom, but none systematically addresses safety risks. In telecom specifically,~\cite{fp3} highlight the need for system-level safety and runtime assurance in GenAI-based multi-agent systems, urging development of compositional explainability and robust assurances across heterogeneous models. 

Freitas et al.~\cite{fp1} survey safety risks such as hallucination, misinformation, and disinformation, auditing commercial chatbots in a mental health context and exposing risky response patterns. Wang et al.~\cite{t4} propose a proactive multi-agent reinforcement learning (MARL) framework integrated with GenAI to handle high dimensionality and non-stationarity, highlighting safety concerns unique to MARL environments. Bellay et al.~\cite{t5} further outline risks for multi-agent systems, including instability, recursive behavior, and incomplete contextual awareness, and propose system-level strategies to increase control over the generative context. These studies illustrate growing recognition of GenAI safety risks, but they primarily focus on single-agent contexts or general MARL environments, rather than domain-specific, high-stakes infrastructures like telecom. 

Wang et al.~\cite{wang2024towards} propose a benchmark for assessing social interaction capabilities of generative agents through objective action-level metrics. Samuel et al.~\cite{samuel2024personagym} introduce PersonaGym, a dynamic evaluation framework for agents, alongside an automatic, human-aligned metric for measuring persona adherence. Mao et al.~\cite{mao2023alympics} present a simulation framework leveraging LLM agents for game theory research in abstract socioeconomic scenarios. These frameworks advance the evaluation of language-agent interactions, but their emphasis is on social dynamics and persona alignment rather than safety in mission-critical domains. 
In summary, prior studies establish the feasibility of GMAS in telecom and highlights safety challenges in other domains, but there remains a gap in domain-specific frameworks that systematically evaluate risks such as miscoordination and information asymmetry in telecom multi-agent settings. Our work addresses this gap by proposing an autonomous evaluation framework that integrates safety, technical correctness, and feasibility into the assessment of GMAS for telecom networks.

\section{System Design}\label{sec:sd}
This section presents the proposed GMAS and the safety evaluation framework for safe agent governance and execution. The framework supports deployment across multiple telecom layers (e.g., edge, RIC, and core orchestration), where agents may operate with differential visibility or symmetric context. 
As shown in Fig.~\ref{fig:architecture}, the GMAS is organized into three functional subsystems and supporting modules. The system is triggered by a systematically generated question, framing a problem or request in the network, which is first received by the Coordinator Agent to initiate orchestration.
\begin{itemize}[left=0pt, itemsep=0pt, labelsep=5pt] 
\item \textbf{Orchestration}: Anchored by the Coordinator Agent, which manages run life cycle, selects solution paths proposed by the Planner Agent, and orchestrates downstream execution. The Planner and Coordinator leverage Tree-of-Thoughts (ToT)~\cite{yao2023tree}, which extends LLM reasoning by exploring multiple solution paths, self-evaluating intermediate steps, and selecting the most promising trajectory. ToT enables both correction of poor initial decisions and robustness through exploration, thereby improving reliability in telecom planning tasks and RAN optimization under uncertain patterns.
\item \textbf{Execution}: Comprises the Resource Allocator Agent, which generates RIC resource allocation plans in pseudo-code, and the Coder Agent, which translates these plans into executable code.
\item \textbf{Analysis}: This involves the Analyzer, which performs static analysis, policy compliance, runtime monitoring, and formal verification, with failed checks triggering targeted refinements. Complementing this, the Safety and Governance module tracks agent runs, interactions, and knowledge asymmetries to detect risks and miscoordination, ensuring safe operation at the system level.
\end{itemize}
\noindent \textbf{Personas}: It define the role and behavioral orientation of each agent in the GMAS ecosystem. Beyond user-facing interactions, personas also shape reasoning. prior work~\cite{samuel2024personagym,mao2023alympics} shows that role-specific personas can improve reasoning compared to zero-shot prompting, though poorly designed personas may introduce bias. In our system, personas are generated automatically using an LLM, guided by each agent’s responsibilities. We evaluate combinations of personas across agents to study how diversity influences individual actions and overall system performance.\\

\noindent \textbf{Knowledge Layer: RAG and GraphRAG} In addition to persona-driven behavior, agent reasoning is refined through domain-grounded knowledge. We integrate both RAG~\cite{lewis2020retrieval} and GraphRAG via a context provider interface: RAG retrieves unstructured documents, while GraphRAG supplies structured relational context from a telecom-specific knowledge graph. The Planner leverages GraphRAG (standards, research papers) to generate multi-step solution paths, which the Coordinator selects for execution. Downstream, the Resource Allocator applies graph-derived O-RAN specifications for compliant strategies, and the Coder retrieves from the FlexRIC codebase to produce executable, context-aware code. \\

\noindent \textbf{Memory and State Management}: 
We employ \textbf{episodic memory}, which captures the past actions of each agent. This memory guides subsequent decisions by allowing agents to recall how earlier tasks were approached and refined, thereby reducing repetition and improving task efficiency. The \emph{Memory Manager} maintains per-agent trajectories, including inputs, thought summaries, and refinement reasons. State artifacts—such as inputs, outputs, embeddings, metrics, and errors—are persisted as JSON and can be aggregated into CSVs for analysis or dashboards. This structure ensures that agents operate not only with domain knowledge but also with continuity of experience and traceable decision-making.\\

\noindent \textbf{Analyzer and Safety Pipeline}
The Analyzer Agent performs multi-dimension validation across:(i)\textbf{ Static checks}: e.g., Pylint, Bandit for Python; Clang-based tools for C++. (ii) \textbf{ Policy compliance}: AST-based enforcement of constraints such as forbidden APIs, and action conflicts. (iii) \textbf{Runtime monitoring}: sandboxed execution and KPI tracking (e.g., throughput, latency). (iv)\textbf{ Formal-lite verification}: lightweight guarantees via type hints, exception handling, and deterministic execution patterns. If the checks fail, the refinement loop triggers targeted re-execution from the relevant agent. The safety pipeline is responsible for analyzing and monitoring potential risk pathways to ensure safe operation of the GMAS. Distinct from the Analyser Agent, the pipeline traces all runs, actions, and interactions between agents, examining how system variables including personas and asymmetric knowledge may contribute to the miscoordination risks and unsafe behaviors. The safety pipeline operates across three stages: (i) \textit{Design-time:} The alignment checks between the solution paths generated by the planer and the Coordinator selection of most promising plan, static and policy checks are applied before execution. (ii) \textit{Execution-time:} Sandboxed runs and testbed simulations validate runtime behavior and operational compliance. (iii) \textit{Post-execution:} Embedding-driven drift detection, and operational metrics (e.g., contextual conflict rate, coordination overhead) assess alignment, consistency, and overall system health.\\

\noindent \textbf{Extensibility and Deployment}: 
The framework is highly modular and extensible, allowing new domain agents, static or policy checkers, runtime harnesses, or evaluation modules to be easily integrated. Personas can be customized per agent to modulate behavior, while governance knobs allow thresholds for drift, alignment, and refinement depth to be configured.  Additionally, the framework can operate autonomously: given a domain specification, it can automatically generate relevant questions or tasks for the agents to address, without manual intervention. This capability enables rapid adaptation to new telecom scenarios or other domains, supporting end-to-end evaluation and refinement cycles. Deployment options range from local sandboxed environments for testing, to lab or testbed setups for KPI-driven validation, and ultimately to production monitoring with strict policy enforcement and human-in-the-loop oversight.
\begin{figure*}[!htbp]
    \centering
    \includegraphics[clip, trim=0cm 2cm 3cm 2cm,width=\textwidth]{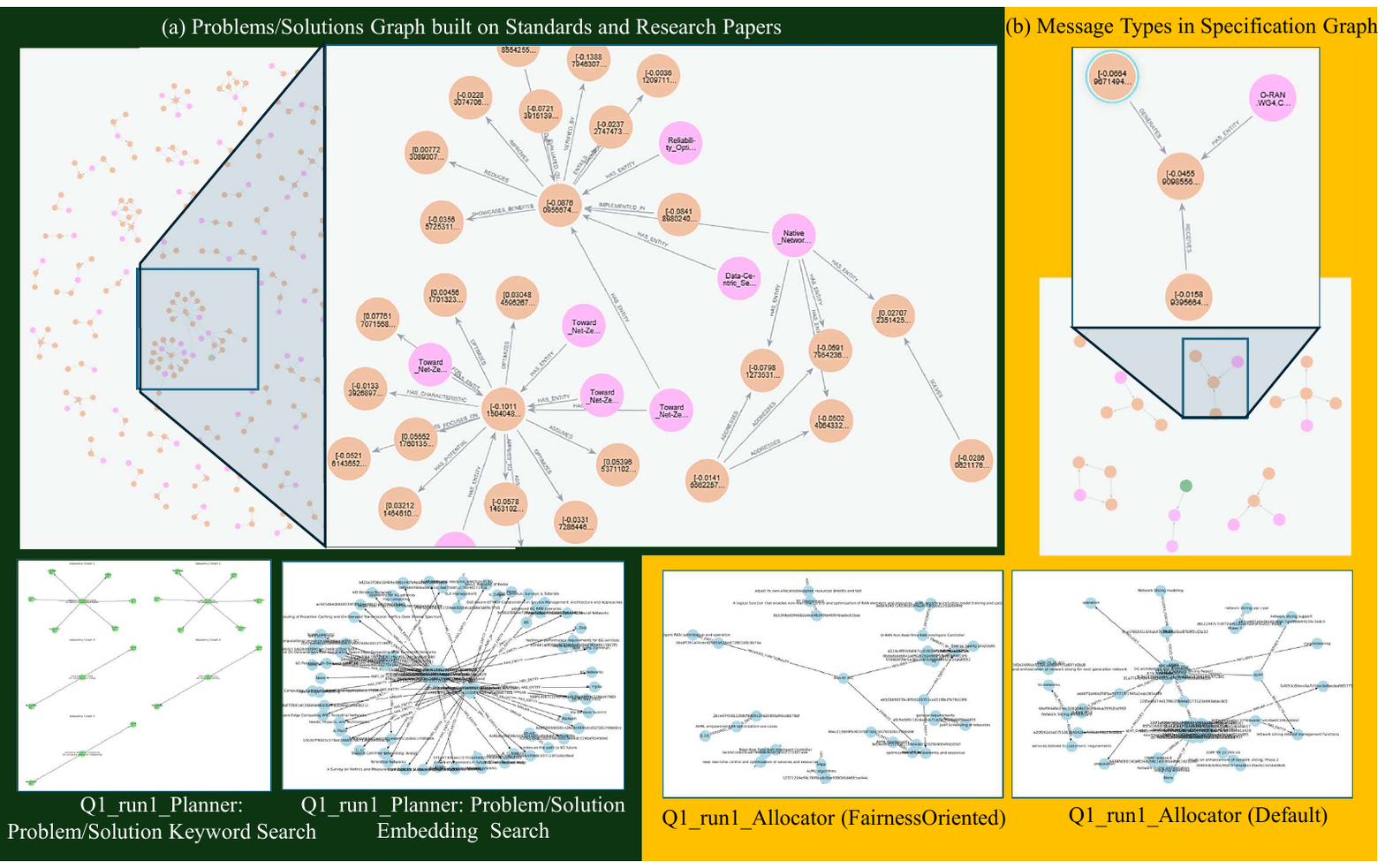}
    \caption{(a) Knowledge graph constructed from O-RAN standards and research papers, supporting Planner Agent in problem-solution searches via keyword and embedding retrieval. (b) Knowledge graph derived from O-RAN specifications, enabling the Allocator Agent to generate pseudocode through embedding-based retrieval.}
    \label{fig:rag}
\end{figure*}
\section{Evaluation}\label{sec:eva}
This section evaluates the performance and safety of the proposed framework for GMAS in the telecom domain. Due to space constraints, we report a subset of the results, while the full dataset and additional analyses are available in our GitHub repository.
The framework employs two complementary categories of metrics. \textit{Analyzer Agent metrics} evaluate the technical quality and compliance of agents output across the dimensions described in Section~\ref{sec:sd}. These dimensions are aggregated into an overall \textit{Analyzer Penalty}. In parallel, \textit{Safety Risk metrics} capture system-level risks across interacting agents. Agents consistency measures the consistency between Resource Allocator and Coder intent and execution, and cross-run embedding distance quantifies semantic drift across runs. Together, these metrics provide a multi-dimensional view of both agent-level reliability and system-level safety.
We evaluate the framework on \textbf{five questions} across \textbf{32 persona sets} ($2^5$ combinations), using input data from the \textbf{O-RAN domain}. The system is powered by \textbf{GPT-3.5-turbo} as the language model and \textbf{all-MiniLM-L6-v2} as the embedding model. We generated the evaluation question set using an LLM-based generator (gpt-3.5-turbo, temperature 0.5) to produce realistic, engineering-oriented questions in the O-RAN domain, focusing on resource allocation and network optimization. 
\begin{figure*}[!htbp]
    \centering
    \includegraphics[clip, trim=0cm 0.3cm 0cm 0cm,width=\textwidth]{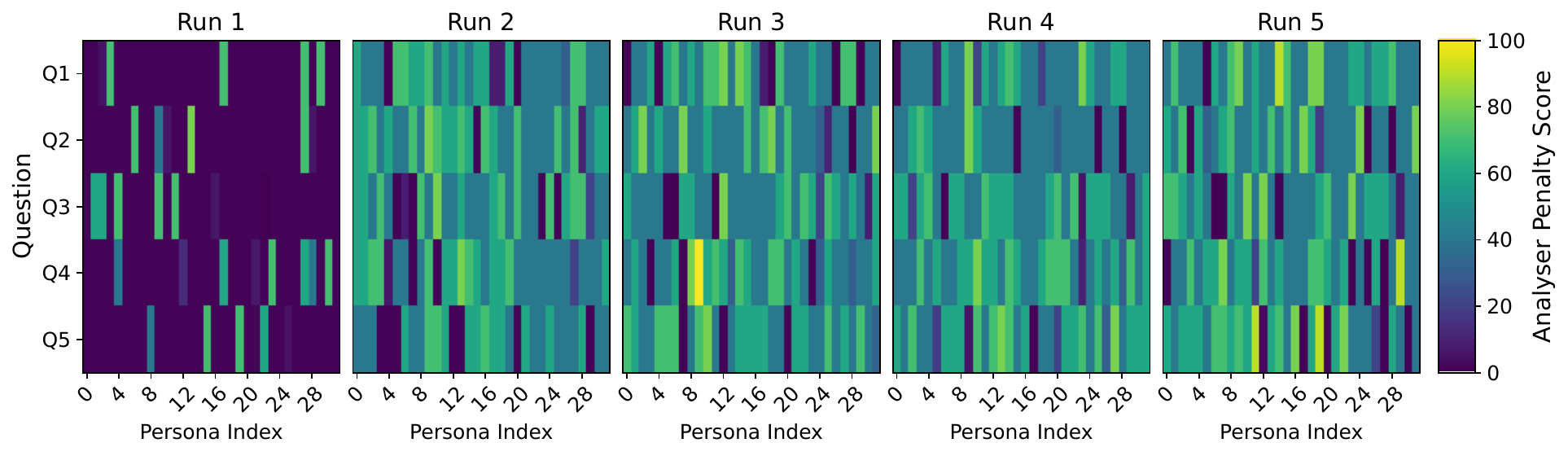}
    \caption{Analyzer Agent penalty score across five runs}
    \label{fig:penalty}
\end{figure*}

\subsection{Analyzer Penalty}
Fig.~\ref{fig:penalty} shows analyzer penalty score across five runs for 32 persona sets. Lower scores indicate more severe issues detected by the Analyzer. Performance improves markedly after the first run: Run 1 starts weak, with a mean score of 10.17 and median of 1.0, suggesting wide-spread alignment and quality issues. From Run 2 onward, the mean rises to around 46–48, and the median stabilizes at 40, indicating that most question-persona combinations achieve moderate-to-high alignment. A small number of persona sets reach scores near 100 suggesting that certain persona sets can yield highly reliavble outputs. In Run 1, 45 out of 160 combinations had the lowest score of 5, concentrated within nine persona sets (28\% of all persona sets). These largely involved “StrictAssessor” persona for the Analyzer or “Default”/“Minimalist” for other agents, indicating elevated safety concerns. Conversely, four persona sets (12.5\%) achieved the highest scores (above 70), often featuring “CreativeThinker” and “FairnessOriented”, which appear to promote more robust and safer outputs.
\par By Run 5, performance consolidates: the three persona sets with the highest averages (60–63) combine “CreativeThinker” and “FairnessOriented” roles in Planner and Resource Allocator with stricter Analyzer settings, yielding the fewest and least severe issues. In contrast, the lowest-scoring sets (28–32) clustered around “StrictAssessor” in Analyzer and “Minimalist” in Coder, suggesting that overly rigid or underspecified personas hindered improvement. Overall, these results highlight substantial variability across personas but also demonstrate the framework’s capacity to surface persona-driven risks, trace their impact on system safety, and identify configurations that foster safer coordination.

\subsection{Agents Consistency}
\begin{figure}[!htbp]
    \centering
    \includegraphics[clip, trim=0cm 0.3cm 3.5cm 0cm,width=\columnwidth]{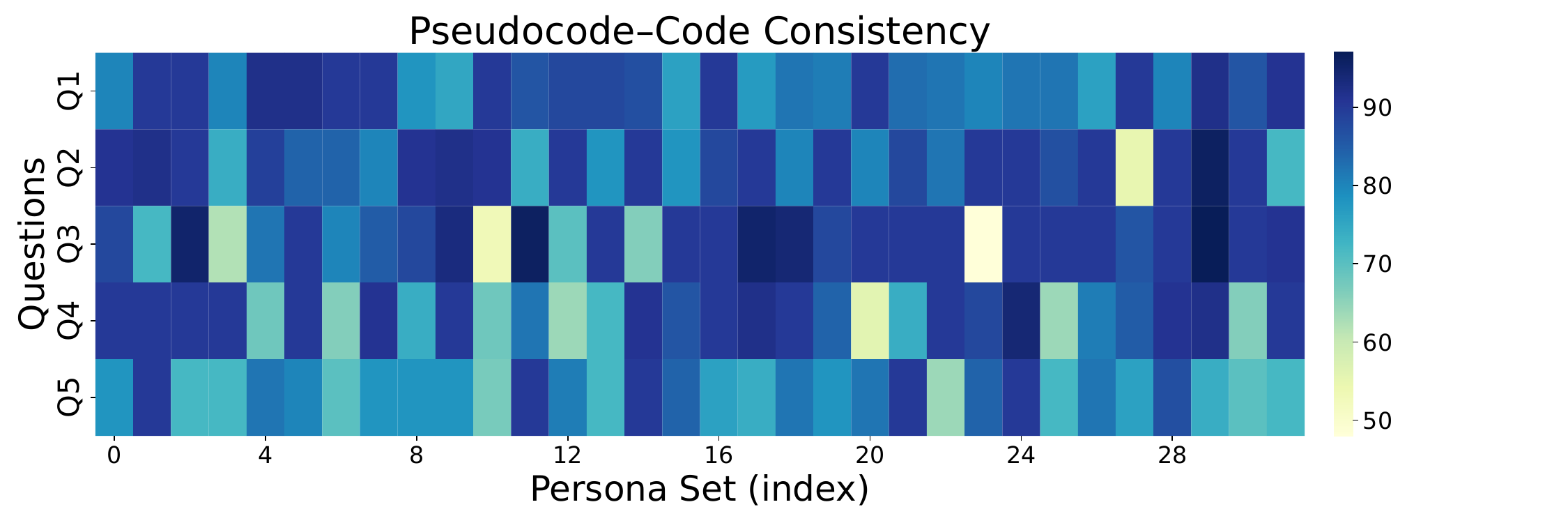}
    \caption{Consistency scores between pseudo-code from Allocator and executable code from Coder.}
    \label{fig:rescod}
\end{figure}

Fig.~\ref{fig:rescod} shows consistency scores between pseudo-code generated by the Resource Allocator and the executable code produced by the CodeAgent. Coding style emerges as the dominant factor: Minimalist Coders consistently achieve stronger alignment, averaging 81–85 with standard deviations of 10–17, while Default Coders exhibit wider variability (73–87, with deviations up to 23). For instance, the configuration \textit{Analyzer:Default, Coder:Minimalist, Coordinator:Default, Planner:CreativeThinker, Allocator:FairnessOriented} achieved 85.6, compared to 73.8 for its Default Coder counterpart. Planner personas such as CreativeThinker improve alignment marginally (2–5 points), but Allocator–Coder consistency is primarily driven by coding style.
These findings underscore that persona design in the coding role directly influences system stability: Minimalist Coders yield more predictable Allocator–Coder translations, while Default configurations introduce higher risk of divergence. The framework thus surfaces how localized persona choices propagate into system-level coordination, an essential property for ensuring safe execution in telecom workflows.

\subsection{Cross-Run Embedding Distance}
\begin{figure}[!htbp]
    \centering
    \includegraphics[clip, trim=0cm 0.5cm 0cm 0.4cm,width=\columnwidth]{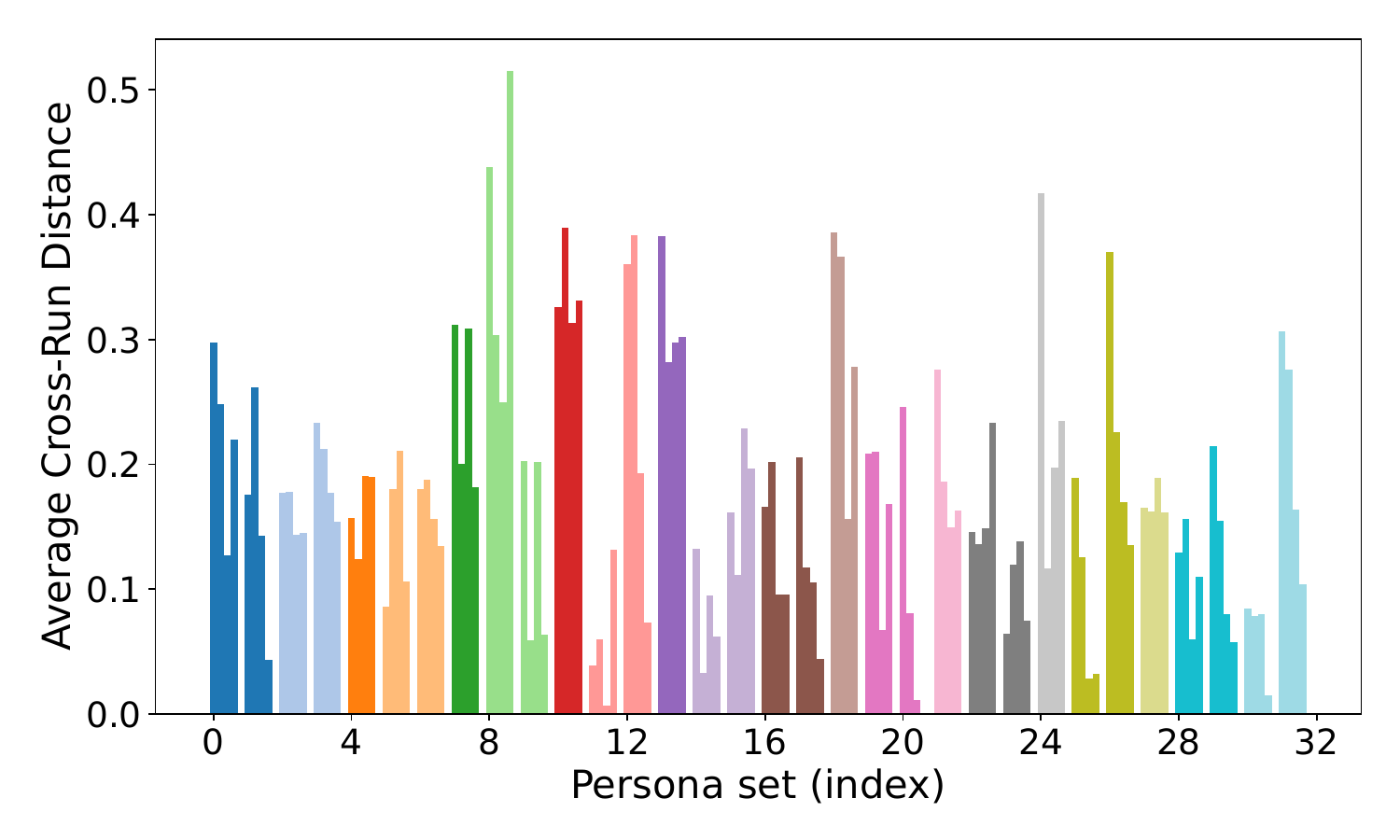}
    \caption{Aggregate cross-run distances for each persona group (Coder Agent). Bars show the mean distance between runs across all questions, with each color representing a persona group.}
    \label{fig:coderembedun}
\end{figure}

Fig.~\ref{fig:coderembedun} shows the average embedding distance between consecutive runs for the Coder, averaged across all questions. On average, distances decrease from run1 to run2 (0.226) to run4 to run5 (0.150), indicating that outputs become more consistent over successive runs. Standard deviations remain relatively high compared to the means, reflecting variability among specific outputs in each transition. 
Distances vary substantially by persona. Combinations such as Minimalist Coder, Default Coordinator, CreativeThinker Planer, and Default Resource Allocator yield high average distances (~0.38), reflecting greater variability, while others, such as Minimalist Coder, Default Coordinator, Default Planer, FairnessOriented Resource Allocator achieve very low averages (~0.06), indicating stability. Overall, coding style, planning approach, and reasoning orientation strongly influence output consistency of agent behavior, with some combinations producing more reproducible results and others yielding more diverse or exploratory outputs.
The Minimalist Coder leads to the highest variability (up to 0.37665), whereas FairnessOriented Allocator reduces distances to as low as 0.05951. Default Coder show moderate variability (0.1457 to 0.2506), while strategic Coordinators slightly lower variability. Overall, Coder and Resource Allocator choices have the largest impact on embedding consistency, with Planner and Coordinator exerting secondary but still noticeable effects in modulating cross-run differences.
Embedding distance analysis shows progressive convergence across runs, but variability is highly sensitive to Coder and Resource Allocator personas, making them the most critical levers for stability.

\section{Conclusion}\label{sec:con}
This paper introduced a GMAS for safety evaluation in the telecom domain, with a focus on O-RAN and RIC settings. The framework integrates agent-level checks and system-level safety metrics, enabling fine-grained evaluation of technical correctness, policy compliance, runtime behavior, and formal guarantees, alongside system-oriented measures such as decision alignment, semantic drift, and coordination load. Through experiments across multiple persona configurations and refining runs, we observed that agent outputs converge over time, with analyzer penalties and embedding distances demonstrating progressive improvements in safety and consistency. The results also highlighted the strong influence of persona design on output stability and reliability. By bridging the gap between domain-specific deployment prospects and responsible AI evaluation, this work contributes a structured and extensible methodology for auditing and governing GMAS in telecom. Future work will extend the framework to larger-scale testbeds, integrate human-in-the-loop safety governance, and explore adaptive policies for dynamic persona adjustment under real-world operating conditions.

\section{Acknowledgment}
This work was supported by EPSRC grants EP/Y037421/1 (CHEDDAR) and UKRI851 on AI-agent cooperation for telecom safety and governance.

\bibliographystyle{unsrt}
\bibliography{main}


\end{document}